\documentclass[preprint,nofootinbib,superscriptaddress,showpacs]{revtex4}

\usepackage{amssymb}
\usepackage{amsmath}
\usepackage{verbatim}
\usepackage{mathrsfs}
\usepackage{amsfonts}
\usepackage{latexsym}
\usepackage{epsfig}
\usepackage{color}
\usepackage{graphicx,subfigure}
\usepackage{units}
\usepackage{slashbox}
\usepackage{natbib}

\begin{document}

%%%%%%%%%%%%%%%%%
%%%   TITLE   %%%
%%%%%%%%%%%%%%%%%
 
\title{Asymmetric condensed dark matter}

\author{Anthony Aguirre}
%\email[]{}
\affiliation{Santa Cruz Institute for Particle Physics and Department of Physics,
University of California, Santa Cruz, CA, 95064, USA}

\author{Alberto Diez-Tejedor}
%\email[]{alberto.diez@fisica.ugto.mx}
\affiliation{Santa Cruz Institute for Particle Physics and Department of Physics,
University of California, Santa Cruz, CA, 95064, USA}
\affiliation{Departamento de F\'isica, Divisi\'on de Ciencias e Ingenier\'ias,
Campus Le\'on, Universidad de Guanajuato, Le\'on 37150, M\'exico}

\date{\today}

%%%%%%%%%%%%%%%%%%%%
%%%   ABSTRACT   %%%
%%%%%%%%%%%%%%%%%%%%

\begin{abstract} 

We explore the viability of a boson dark matter candidate with an asymmetry between the number densities of particles and antiparticles. 
A simple thermal field theory analysis confirms that, under certain general conditions, this component 
would develop a Bose-Einstein condensate in the early universe that, for appropriate model parameters,
could survive the ensuing cosmological evolution until now. 
The condensation of a dark matter component in equilibrium with the thermal plasma is a relativistic process, 
hence the amount of matter dictated by the charge asymmetry is complemented 
by a hot relic density frozen out at the time of decoupling. Contrary to the case of ordinary WIMPs, dark matter 
particles in a condensate must be lighter than a few tens of eV so that 
the density from thermal relics is not too large. Big-Bang nucleosynthesis 
constrains the temperature of decoupling to the scale of the QCD phase transition or above. 
This requires large dark matter-to-photon ratios and very
weak interactions with standard model particles.

\end{abstract}

%%%%%%%%%%%%%%%%
%%%   PACS   %%%
%%%%%%%%%%%%%%%%


%%%%%%%%%%%%%%%%%%%%%%
%%%   MAKE TITLE   %%%
%%%%%%%%%%%%%%%%%%%%%%

\maketitle

%%%%%%%%%%%%%%%%%%%%%%%%
%%%   INTRODUCTION   %%%
%%%%%%%%%%%%%%%%%%%%%%%%

\section{Introduction}
\label{sec:introduction}

The idea of a weakly interacting massive particle (WIMP) has guided the research in dark matter (DM) over the last three decades~\cite{WIMPs, ProfumoTASI, KolbTurner}.
In the standard cosmological scenario WIMPs are produced thermally in the early universe, with an abundance of particles today fixed at the time of thermal 
(chemical) decoupling. The energy density in WIMPs is mainly determined by their self-annihilation rates into the standard model sector. 
For particles with masses and couplings at the electroweak scale the energy density in thermal relics roughly coincides with the known abundance of DM. 
Particles with these properties appear in e.g. supersymmetric extensions to the standard model, and this coincidence is usually known as the 
``WIMP miracle."

However, there are alternatives that deserve to be considered seriously. We know that the amount of baryons in the universe is not determined at thermal decoupling, 
but by an asymmetry between the number densities of particles and antiparticles. Something similar could have also happened to DM~\cite{asymmetric}. 
If in addition to the existence of a nonvanishing conserved charge, DM is described in terms of a boson particle that was in thermal 
equilibrium with the constituents of the standard model in the early universe, it is natural to think that the zero-mode could have developed a 
nonvanishing expectation value, leading to the appearance of a cosmological Bose-Einstein condensate (BEC)~\cite{Luis, luke, Li}. 
We call this scenario {\it asymmetric condensed dark matter}. Other proposals with a similar spirit have been previously considered under many 
different names~\cite{Qballs,fuzzy,FluidDM,Goodman,Sahni,luis-tona,Arbey,Harko,fabio,
Barranco,chavanis,shapiro,Lora,Matos,diez,Tom,marsh1,guzman,khoury}. 
See also Refs.~\cite{sikivie, sikivie2, Berges, Davidson, Chanda} 
for similar ideas in the context of the QCD axion, where unlike 
the case considered here DM never thermalizes with the standard model particles in the early universe.

In this paper we identify four conditions that, together, guarantee the appearance of a cosmological thermal condensate
of particles that could match the totality of the cold dark matter (CDM) today: 
\begin{enumerate}
\item DM is described in terms of a boson field with a conserved current, 
\item There is a nonvanishing total boson charge with the right amount in the universe, 
\item The boson was in thermal (kinetic and chemical) equilibrium with the particles in the standard model,  
\item Thermal decoupling took place at an energy scale below the critical temperature of condensation. 
\end{enumerate}
Under these assumptions we conclude that a thermal BEC can only emerge from a relativistic process, and that the particles composing the condensate 
should be much lighter than standard WIMPs. This requires DM-to-photon ratios that are at least eight orders of magnitude larger than in the case of 
baryons, a point that deserves further attention. It is this large asymmetry what 
makes it possible that light particles are nonrelativistic in spite of being in thermal equilibrium in the early universe.
The interaction of the DM bosons with the standard model particles must be very weak, even more than the weak nuclear force, although strong enough 
to reach early thermal equilibrium. We speculate about the possibility that the gravitational interaction alone could bring the DM bosons into thermal 
equilibrium with standard model particles, but the naive analysis we present in this paper deserves a further study in terms of the Boltzmann equation.
Figure~\ref{fig:cosmological evolution} in Section~\ref{sec:BECDM} summarizes the main features of this proposal, that remain valid if the condensate contributes only to a partial 
fraction of the total CDM in the universe. 

Note that the four conditions above lead to a macroscopic excitation of the zero-mode that in the present universe is not precisely a thermal 
BEC by the standard definition, i.e. it has not been in thermal equilibrium
since DM decoupling. We will nonetheless refer to this state as the {\it condensate} in this paper. 
We should emphasize that the description of a thermal condensate we present here is only valid as long as
the expansion rate of the universe is lower than the characteristic frequency of oscillation of the zero-mode.
Furthermore, in order to not affect the abundance of light elements predicted in the standard cosmological scenario
the DM particles should be already decoupled from the thermal plasma before the outset of Big-Bang nucleosynthesis (BBN), 
and then we cannot extrapolate the results in this work to candidates lighter than around $m\sim 10^{-14}\,\textrm{eV}$. 
Further constraints on the temperature of decoupling coming from the effective number of extra neutrino species increase this value to 
$10^{-12}\,\textrm{eV}$. Particles with a mass below this number require a more detailed analysis that we leave for another paper.
We also stress that our motivation here is purely phenomenological, and we do not develop  any specific boson candidate (a possible particle realization 
is discussed in e.g. Ref.~\cite{axiverse}, see also~\cite{MarshReview}). Rather, we assume an asymmetry and an early stage of thermal equilibrium, and explore the consequences 
of demanding a late-time coherent behavior on the model. 

The paper is organized as follows. In Section~\ref{sec:formalism} we review some basics on quantum fields in curved spaces, and frame 
the idea of a cosmological thermal condensate of scalar particles within our 
current understanding of physical theory. Then in Section~\ref{sec:BECDM} we use some basic cosmological observations to constraint the parameters of 
the boson as a model of DM. Finally, we conclude in Section~\ref{sec:discussion} with a brief discussion of the main results of this paper.

%%%%%%%%%%%%%%%%%%%%%%%%
%%%   FORMALISM   %%%
%%%%%%%%%%%%%%%%%%%%%%%%

\section{Cosmological gravitating bosons in thermal equilibrium}\label{sec:formalism}

Let us first consider the general problem of a cosmological gravitating condensate. In this paper we assume that the spacetime metric is a classical 
object, but matter is described in the language of quantum field theory. This picture is expected to be appropriate as long as we are only interested 
in describing physical processes taking place well outside the Planck regime, with the classical spacetime and the quantum matter fields connected 
through the semiclassical Einstein field equations, $G_{\mu\nu}=8\pi G \langle\hat{T}_{\mu\nu}\rangle$. 
Here $G_{\mu\nu}$ is the Einstein tensor, $G$ denotes Newton's gravitational constant, and
$\langle\hat{T}_{\mu\nu}\rangle$ is the expectation value of the energy-momentum tensor operator, discussed further below.

In this section we consider the case of a cosmological gas of bosons in thermal equilibrium.  
At this point we do not need to specify how this thermal regime could have been achieved (and even supported) 
in practice. The process of DM decoupling, essential for any sensible condensate model, is discussed next in Section~\ref{sec:BECDM}.

\subsection{Quantum field theory in curved spacetimes: a minimal review}\label{sec:review}

We restrict our attention to the case of a massive, complex scalar field with an internal $U(1)$ global symmetry and no self-interactions. 
Associated with the global symmetry there is a conserved current, and then (crucially) the possibility to have a non vanishing conserved charge. 
The generalization to more involved fields and/or symmetries is possible, but we do not consider it here. 
Classically, the scalar field $\varphi(x)$ satisfies the Klein-Gordon equation,
\begin{equation}\label{KG}
(\Box-m^2)\varphi(x) = 0 \,,
\end{equation}
with $\Box \equiv g^{\mu\nu}\nabla_{\mu}\nabla_{\nu}$ the d'Alembert operator in four dimensions and $x$ a generic point in the spacetime manifold. 
A minimal coupling between the scalar field and the gravitational interaction has been considered here. We are using units with $c = \hbar = 1$, 
where the scalar field $\varphi(x)$ and the mass parameter $m$ have dimensions of energy.

Given two solutions $\varphi_{1}(x)$ and $\varphi_{2}(x)$ to the classical equation of motion, Eq.~(\ref{KG}), we can define the symplectic 
(or Klein-Gordon) scalar product
\begin{equation}\label{scalar}
 (\varphi_1,\varphi_2)%_{\textrm{Sympl}}
 \equiv - i \int_{\Sigma}[\varphi_1(\partial_{\mu}\varphi_2^*)-
 (\partial_{\mu}\varphi_1)\varphi_2^*]\,n^{\mu} d\Sigma\, .
\end{equation}
Here $n^{\mu}$ is a timelike, future-directed, normalized four-vector orthogonal to the three-dimensional Cauchy hypersurface $\Sigma$, and $d\Sigma = \sqrt{\gamma}\,d^3 x$ 
is its volume element. The function $\varphi_2^*(x)$ in Eq.~(\ref{scalar}) denotes the complex conjugate of $\varphi_2(x)$. By construction the Klein-Gordon scalar product  
does not depend on the Cauchy hypersurface $\Sigma$, and thus is preserved with time evolution.

At the quantum level the state of the system is described in terms of a vector in a Hilbert space, with the
field, $\varphi(x)$, and its conjugate momentum, $\pi(x)=\sqrt{\gamma}(n^{\mu}\partial_{\mu}\varphi^*(x))$, promoted to operators acting 
on the elements of this space~\cite{BirrellDavies,Wald,Mukhanov,Parker}. In this paper we follow the canonical approach and impose the standard equal-time 
commutation relations
\begin{equation}\label{commutators}
 [\hat{\varphi}(t,\vec{x}),\hat{\pi}(t,\vec{y})]=i\delta(\vec{x}-\vec{y})\, , \quad
 [\hat{\varphi}(t,\vec{x}),\hat{\varphi}(t,\vec{y})]=[\hat{\pi}(t,\vec{x}),\hat{\pi}(t,\vec{y})]=0\, .
\end{equation}
We are working in the Heisenberg picture, where the state vectors remain fixed and only the field operators evolve in time.
As usual we can decompose 
$\hat{\varphi}(x)$ in terms of the time-independent creation and annihilation operators,
\begin{equation}\label{expansion}
 \hat{\varphi}(x)=\sum_{i}(\hat{a}_{i} u_i (x)+\hat{b}^{\dagger}_{i} u_i^* (x))\, ,
\end{equation}
with $u_i (x)$ a complete set of normal modes, each solving Eq.~(\ref{KG}),
that are orthonormal with respect to the symplectic product: 
\begin{equation}\label{normalization}
(u_i,u_j) = \delta_{ij}\,,\quad (u^*_i,u^*_j) = -\delta_{ij}\,,\quad (u_i,u^*_j) = 0\,.
\end{equation}
Here the letters $i$ and $j$ are just labels for the mode-functions, and not spacetime indexes.
The normal modes $u_i(x)$ in Eq.~(\ref{expansion}) have dimensions of energy, with 
the creation, $\hat{a}_i^{\dagger}$,  $\hat{b}_i^{\dagger}$, and annihilation, $\hat{a}_i$, $\hat{b}_i$, operators dimensionless. 
All these conventions guarantee the standard commutation relations 
$[\hat{a}_i,\hat{a}_j^{\dagger}]=[\hat{b}_i,\hat{b}_j^{\dagger}]=\delta_{ij}$, with all other 
combinations zero. Note that in the continuum limit the sum in Eq.~(\ref{expansion}) should be replaced by an integral, and then the Kronecker delta in 
Eq.~(\ref{normalization}) by a Dirac delta function, but the construction is similar. 

The vacuum is defined to be the state for which $\hat{a}_i\vert 0 \rangle = \hat{b}_i\vert 0 \rangle =0$ for all $i$.
The Hilbert space can then be constructed ({\it \`a la} Fock) by successive applications of creation operators on the vacuum state.
Incidentally, the elements of this 
construction, $| N_i^a,\ldots, N_i^b,\ldots\rangle$, are the eigenstates of the particle ($\hat{N}_i^a= \hat{a}_i^{\dagger}\hat{a}_i$) and antiparticle ($\hat{N}_i^b =
\hat{b}_i^{\dagger}\hat{b}_i$) number operators.

If the state of the matter fields is described by a vector $\vert\psi\rangle$ in Hilbert space (i.e. a pure state), then $\langle\hat{\mathcal{O}}(x)\rangle =
\langle\psi\vert\hat{\mathcal{O}}(x)\vert\psi\rangle$, with $\hat{\mathcal{O}}(x)$ denoting a generic observable in the theory. However, 
we wish to describe %charged ---with respect to the dark $U(1)$ symmetry group--- 
boson particles at finite temperature,
$k_B T=1/\beta$, %that are charged with respect to an internal dark $U(1)$ symmetry group. Here 
where $k_B=1$ is the Boltzmann constant. In this latter case we should replace the standard expectation value 
$\langle\hat{\mathcal{O}}(x)\rangle$ above by~\cite{Fetter,kapusta}
\begin{equation}\label{eq.thermal.exp}
 \langle \hat{\mathcal{O}} (x) \rangle_{\beta}=
 \frac{\textrm{Tr}[e^{-\beta(\hat{H}-\mu\hat{Q})}\hat{\mathcal{O}}(x)]}{\textrm{Tr}[e^{-\beta(\hat{H}-\mu\hat{Q})}]} \, .
\end{equation}
Here $\textrm{Tr}[e^{-\beta(\hat{H}-\mu\hat{Q})}]$, the partition function, represents the trace of the density matrix in the grand canonical ensemble, with $\hat{H}$ 
the Hamiltonian operator for the scalar field and $\mu$ the chemical potential associated to the conserved charge 
$\hat{Q}=-i\int[\hat{\varphi}\hat{\pi}-\hat{\pi}^\dagger\hat{\varphi}^\dagger]d^3x$.
This charge essentially represents the difference between the number of particles and antiparticles in the configuration.

$\vspace{0.01cm}$

\centerline{*\quad *\quad *}

$\vspace{0.01cm}$

Next we derive a number of cosmologically-relevant expressions for this model. Note that many of these equations would also apply if there are
no conserved charges in the dark sector, $\mu=0$, in which case it is not possible to develop a thermal condensate of DM particles.
%If there were no conserved charges we would simply write $\mu=0$ and all the expressions below would apply. Note however that it is not possible to 
%develop a thermal Bose-Einstein condensate without a nonvanishing conserved charge in the configuration. 
For a matter component described in terms of 
a real field particles and antiparticles coincide, and we cannot define properly a conserved charge operator $\hat{Q}$ in the system. 
However, if this field were not coupled to the standard model of particle physics (or if the interactions with this sector were weak enough), 
it could be possible to reach a diluted state of nonrelativistic kinetic equilibrium where, at the effective level, the number of particles were conserved. 
Note however that this state would not be in chemical equilibrium with the cosmological thermal plasma, as we assume in this paper. 

It has been recently proposed that this could happen for the QCD axions~\cite{sikivie, sikivie2}, where the particles are generated initially 
in a low momentum state through a misalignment of the vacuum in the early universe~\cite{Dine,Wilczek,Gondolo1,Gondolo2,Wantz}, 
although there is still some debate in the literature~\cite{Berges, Davidson, Chanda}. In any case, the temperature in axions would not coincide 
with the temperature of the standard model particles, and then the results in this paper do not necessarily apply there.

%%%%%%%%%%%%%%%%%%%%%%%%%%%%%%%%%
%%%   COSMOLOGY	   %%%
%%%%%%%%%%%%%%%%%%%%%%%%%%%%%%%%%

\subsection{Cosmological considerations}
\label{sec.cosmology}

In this section we apply a simple analysis in field theory to describe an asymmetric thermal gas of bosons in an expanding universe. We  
concentrate on the appearance of a cosmological condensate in the limiting case of a relativistic regime. 
As we will find next in Section~\ref{sec:BECDM} the condensation of a DM component in equilibrium with the thermal plasma is 
necessarily a relativistic process, and then a relativistic analysis is mandatory for the purposes of this paper.

In absence of gravity, the emergence of a thermal condensate has been extensively considered in the 
literature~\cite{Dalfovo, Pitaevskii, Pethick, Griffin, Huang, Haber, others, grether}. 
Let us now focus on the case in which the boson particles live on a spatially flat, homogeneous and isotropic universe. 
Under this assumption the spacetime metric can be written in the form
\begin{equation}\label{eq.FRW}
 ds^2 = -dt^2 + a^2(t) (dx^2 +dy^2 +dz^2)\,.
\end{equation}
Here $t$ is the cosmological time, $a(t)$ the scale factor, and $(x,y,z)$ a set of comoving spatial coordinates.
Without any loss of generality we can choose $a=1$ today, when comoving and physical quantities coincide.
To proceed, we require a set of normal modes adapted to the cosmological background.
Because there is no timelike Killing vector field
associated to the line-element in Eq.~(\ref{eq.FRW}), there will not exist a set of normal
modes in the form of ordinary plane-waves, 
and most of the results in flat spacetime do not formally apply.
However, if the expansion rate of the universe is not too large,
$H\ll m$, we can still write
\begin{equation}\label{cosmo.waves}
 u_{\vec{k}}(x)= \frac{1}{\sqrt{2Va^3\omega_k}}\exp\left(-i\int\omega_k dt+i\vec{k}\cdot\vec{x}\right) \,,
\end{equation}
where $H\equiv\dot{a}/a$ is the Hubble parameter, and $\omega_k\equiv\sqrt{m^2+a^{-2}k^2}$ the dispersion relation of the scalar particle. 
We are imposing periodic boundary conditions over a box of {\it comoving} size $L$, with $\vec{k}=(2\pi/L)(n_x,n_y,n_z)$, $V=L^3$, and $n_x,n_y,n_z=0,\pm1,\pm2,\ldots$.
Note that the wavenumber $\vec{k}$ labels the different mode-functions in Eq.~(\ref{cosmo.waves}), and that the mode-energies $\omega_k$ depend only on $k=|\vec{k}|$;
this is a consequence of the symmetries associated to the slices of constant cosmological time $t=\textrm{const}$. 
However, the spacetime background is not static, and then the dispersion relation
is time-dependent. Remember that we are interested in describing a condensate in an {\it infinite} homogeneous and isotropic flat universe,
and then we can switch to the continuum, $\sum_{\vec{k}}=V/(2\pi)^3\int d^3\vec{k}$, if necessary. 

The functions in Eq.~(\ref{cosmo.waves}) satisfy the Klein-Gordon equation, Eq.~(\ref{KG}), and the simplectic
normalization condition, Eq.~(\ref{normalization}), to the second order in an adiabatic WKB approximation, $H/\omega(k)\ll 1$.
Then, as long as the the expansion rate of the universe is lower than the characteristic frequency of oscillation of the zero-mode, $H<m$, the expression
in Eq.~(\ref{cosmo.waves}) describes the whole spectrum of mode-functions to the next-to-leading order in the adiabatic approximation. 
In this paper we will be interested only in a description of the condensate 
to the zeroth order in this series expansion, and then we will not keep factors of $H/\omega(k)$ in  
the final expressions. 

We are mainly interested in obtaining cosmological {\it thermal} expectation values. First, however, we need to clarify some issues 
related with the vacuum of the theory. Formally, the expectation value of a quantity that is quadratic or higher order in field operators 
diverges, even when it is evaluated on the vacuum state. See for instance the Hamiltonian %and the scalar charge 
as a particular realization of a quadratic operator with ultraviolet divergences. We should then use some regularization/renormalization prescription 
in order to obtain, from these quantities, well defined objects when acting on physical states. There are different methods 
to renormalize higher order operators, such as proper-time regularization, dimensional regularization, zeta-function regularization, point-splitting 
regularization, or adiabatic regularization~\cite{Parker}. For the case of a homogeneous and isotropic universe the simplest choice is probably 
adiabatic regularization. This prescription was employed for the first time by Parker in~\cite{Parker2}, and we can use this technique to identify 
sensible vacuum expectation values from ill-defined higher order operators. While important on their own, these vacuum contributions are not 
really relevant for the problem we want to address in this paper, and we will simply omit them in all the expressions below (see for instance 
Refs.~\cite{Fulling, Parker3} for a renormalized energy-momentum tensor in the context of an expanding homogeneous and isotropic universe). 
In practice, this is essentially equivalent to impose standard normal ordering, e.g.
$:\hat{a}_i\hat{a}_i^{\dagger}:=\hat{a}_i^{\dagger}\hat{a}_i$, $:\hat{b}_i\hat{b}_i^{\dagger}:=\hat{b}_i^{\dagger}\hat{b}_i$, on 
higher order operators, and ignore the divergent zero-point contributions to the expectation values.
With all these assumptions the Hamiltonian and the scalar charge in Eq.~(\ref{eq.thermal.exp}) simplify to
\begin{equation}\label{eq.HQ}
\hat{H} = \sum_{\vec{k}} (\hat{N}_{\vec{k}}^a + \hat{N}_{\vec{k}}^b)\omega_k \,, \quad \hat{Q} = \sum_{\vec{k}} (\hat{N}_{\vec{k}}^a - \hat{N}_{\vec{k}}^b) \,,
\end{equation}
where, as mentioned above, we have omitted nondiagonal terms in the expression for the Hamiltonian operator that are suppressed by factors of 
$H/\omega(k)$ in the series expansion. For completeness, the operator representing the total number of particles (including antiparticles) 
%in the configuration 
is by definition given by $\hat{N}=\sum_i(\hat{N}_i^a + \hat{N}_i^b)$.

Since the Hamiltonian and the scalar charge commute, $[\hat{H},\hat{Q}]=0$, we can choose a basis of the Hilbert space in which both $\hat{H}$ 
and $\hat{Q}$ are diagonal. To the zeroth order in the adiabatic expansion the Hamiltonian and the scalar charge also commute with the 
number operators, $[\hat{N}_{\vec{k}}^a,\hat{H}]=[\hat{N}_{\vec{k}}^b,\hat{H}]=[\hat{N}_{\vec{k}}^a,\hat{Q}]=[\hat{N}_{\vec{k}}^b,\hat{Q}]=0$, 
and we can then use e.g. the basis obtained using the Fock construction, where all the elements have a definite number of particles and antiparticles. 
This choice simplifies the analysis, since only the terms in the diagonal of an operator can contribute to the expectation value
$\langle\hat{\mathcal{O}}(x)\rangle_{\beta}$ in this basis. This is what happens, for instance, in the case of the energy-momentum tensor, 
where in practice we can just forget the other terms in %the expression of 
$\hat{T}_{\mu\nu}(x)$ different to 
$(\hat{N}^a_{\vec{k}}+\hat{N}^b_{\vec{k}})T^{(\vec{k})}_{\mu\nu}(x)$. Here $T^{(\vec{k})}_{\mu\nu}(x)$ is the energy-momentum
tensor corresponding to an excitation in the $\vec{k}$-mode,
\begin{equation}
 T^{(\vec{k})}_{\mu\nu}(x) = \partial_{\mu}u_{\vec{k}} \partial_{\nu}u^*_{\vec{k}} + \partial_{\mu}u^*_{\vec{k}} \partial_{\nu}u_{\vec{k}}
 - g_{\mu\nu}
 \left(\partial_{\sigma}u_{\vec{k}}\partial^{\sigma}u_{\vec{k}}+m^2 u_{\vec{k}} u^*_{\vec{k}}\right)\,,
\end{equation}
and we have again neglected those terms suppressed in the adiabatic expansion.

After some algebra (see for instance Chapter 2 in Ref.~\cite{Fetter} for some details in Minkowski spacetime), 
we obtain the thermal expectation values associated to the 
number density, $n_{\beta}=\langle\hat{N}\rangle_{\beta}/(a^3V)$, charge density, $q_\beta=\langle\hat{Q}\rangle_{\beta}/(a^3V)$, energy density, 
$\rho_\beta=-\langle\hat{T}_0^0(x)\rangle_{\beta}$, and pressure, $p_\beta=\langle\hat{T}_n^n(x)\rangle_{\beta}$ (no sum in $n$), 
of a gas of bosons in a cosmological background,
\begin{subequations}\label{p.rho.sum}
\begin{eqnarray}
 n_\beta &=& (\bar{n}^a_{0}+\bar{n}^b_{0})+\frac{1}{2\pi^2 a^3}\int_0^{\infty} (\bar{N}^a_{k}+\bar{N}^b_{k}) k^2dk\,, \\
 q_\beta &=& (\bar{n}^a_{0}-\bar{n}^b_{0})+\frac{1}{2\pi^2 a^3}\int_0^{\infty} (\bar{N}^a_{k}-\bar{N}^b_{k}) k^2dk\,, \label{scalar.charge.therm}\\
 \rho_\beta &=& (\bar{n}^a_{0}+\bar{n}^b_{0})m+\frac{1}{2\pi^2 a^3}\int_0^{\infty} (\bar{N}^a_{k}+\bar{N}^b_{k}) \omega_{k}k^2dk\,, \label{p.rho.cont.1}\\
 p_\beta &=& \frac{1}{2\pi^2 a^3}\int_0^{\infty} (\bar{N}^a_{k}+\bar{N}^b_{k}) \frac{a^{-2} k^2}{3\omega_{k}}k^2dk\,. \label{p.rho.cont.2}
\end{eqnarray}
\end{subequations}
It is easy to see that  in this coordinate system the other components of the energy-momentum tensor vanish by symmetry considerations. 
Here $\bar{N}_i^a$ and $\bar{N}_i^b$ are the mean occupation number of particles and antiparticles in the different energy levels, respectively,
\begin{equation}\label{eq.thermal.ocupation}
 \bar{N}_k^a = \frac{1}{\exp\left[\beta(\omega_k-\mu)\right]-1}\, , \quad \bar{N}_k^b = \frac{1}{\exp\left[\beta(\omega_k+\mu)\right]-1}\, ,
\end{equation}
with $\bar{n}^a_k=\bar{N}^a_k/(a^3V)$ and $\bar{n}^b_k=\bar{N}^b_k/(a^3V)$. 
Since particles and antiparticles are in chemical equilibrium their corresponding chemical potentials are equal in magnitude but opposite in sign.
%Finally, from Euler's equation we obtain $S_\beta=\beta V(\rho_\beta+p_\beta-\mu q_\beta)$ for the total entropy in the configuration.
From now on we will omit the subindex $\beta$ and the overbars in order to simplify the notation unless this may lead to confusion.

Some comments are in order here. To proceed, we have taken the $L\to\infty $ limit in the size of the comoving normalizing box, and 
approximated sums over energy states by integrals over wavenumbers, $\sum_{\vec{k}}=(V/2\pi)^3\int d^3k$. The integral terms 
coincide with the standard textbook expressions for the number density, charge density, energy density, and pressure 
of a gas of bosons at finite temperature (see e.g. Section~3.3 in Ref.~\cite{KolbTurner} for an example in the cosmology literature). 
However, the density of momentum states $g(k)=(V/2\pi^2)k^2$ vanishes at $k=0$, and then the zero-mode is not contained in the previous integrals. 
That is the reason for which it has been necessary to write $n_0=n^a_{0}+n^b_{0}$, $q_0=n^a_{0}-n^b_{0}$, 
$\rho_0 =(n^a_{0}+n^b_{0})m$, and $p_0=0$ (overbars omitted) explicitly in Eqs.~(\ref{p.rho.sum}).
As we will show next, these additional terms coming from the zero-mode could be relevant (and even dominate) 
at low temperatures if the charge density is different from zero. 

In order to see this note that, since the total charge is conserved in a comoving volume, the expression in Eq.~(\ref{scalar.charge.therm}) 
(implicitly) fixes the value of the chemical potential. As usual, the condition of a positive number of 
particles and antiparticles in the different energy levels demands $|\mu|\le m$.
The mean occupation number is always larger in the zero-mode than in any other 
level, $n_0>n_{k\neq0}$, but usually negligible when compared to the total  
particle density, $n_0\ll n$. 
However, if at a temperature different from zero the chemical potential approaches the critical value at $|\mu|\to m$,  
a macroscopic amount of charge $|q_0| \lesssim |q|$ must be accommodated in the
lowest energy mode, $k=0$. A phase transition takes place, signifying the appearance of a 
BEC~\cite{Dalfovo,Pitaevskii,Pethick,Griffin}. (Note that if we set $\beta\neq 0$, $|\mu|=m$ in Eq.~(\ref{eq.thermal.ocupation}) 
the expression for $n_0$ formally
diverges, but of course in practice the value of the chemical potential is close but not equal to $\pm m$.)

At this point the field operator $\hat{\varphi}(x)$ describing the fundamental degrees of freedom in the gas develops a nonvanishing expectation value,
$\langle\hat{\varphi}(x)\rangle\neq0$, with the mode of lowest energy described in terms of a coherent pure state in Hilbert space,
\begin{equation}\label{coherent.state}
 \hat{a}_0|\psi\rangle = e^{i\theta}\sqrt{Vq_0}|\psi\rangle \,, \quad \hat{a}_{\vec{k}\neq 0}|\psi\rangle = \hat{b}_{\vec{k}}|\psi\rangle = 0 \,,
\end{equation}
i.e. an eigenstate of the annihilation operator $\hat{a}_0$ with eigenvalue $e^{i\theta}\sqrt{Vq_0}$. Here, as usual, we have normalized this state to unity,
$\langle\psi|\psi\rangle=1$, the phase $\theta$ is arbitrary, and (without any loss of generality) we are assuming a universe with more particles 
than antiparticles, $q>0$, i.e. $\mu\in[0,m)$. We can then neglect the mean occupation number of antiparticles in the ground state, 
$n^b_0\ll n^a_0$, and approximate $q_0=n^a_0-n^b_0\approx n^a_0$.\footnote{Note that the results in this paper are not really sensitive 
to the quantum coherence of the state $|\psi\rangle$. If the particles were just accommodated 
in the mode with lowest energy with arbitrary phases (i.e. if this mode were described in terms of an eigenstate of the particle number operator with eigenvalue $Vq_0$, 
$\hat{N}_0^a|\psi\rangle = Vq_0|\psi\rangle$), they would source the same energy momentum-tensor as the state in 
Eq.~(\ref{coherent.state}). However, contrary to the case in Eq.~(\ref{exp.zero.mode}), the expectation value of the field operator would vanish, 
$\langle\hat{\varphi}(x)\rangle=0$. This could have important observational 
consequences when discussing e.g. the direct detection of this matter component, but this is not the subject of this paper.}
The nonvanishing expectation value  
\begin{equation}\label{exp.zero.mode}
 \varphi(x) \equiv \langle \psi| \sum_{\vec{k}}\left(\hat{a}_{\vec{k}}u_{\vec{k}}(x)+\hat{b}^{\dagger}_{\vec{k}}u^*_{\vec{k}}(x) \right) |\psi\rangle = 
 \sqrt{\frac{q_0}{2a^3m}}e^{-i(m t-\theta)} \,
\end{equation}
plays the role of an order parameter for the BEC. Note that this order parameter satisfies the Klein-Gordon equation (in the nonrelativistic limit the Gross-Pitaevskii 
equation for a system of non-interacting particles). We can then describe the condensate in terms of a 
{\it classical field theory} ---classical from the point of view of a field theory, purely quantum from a particle interpretation.
Introducing the expression for the order parameter into the equations for the number density, charge density, energy density, and pressure 
associated to a classical scalar field, we recover the zero-mode contributions 
$n_0= n^a_{0}$, $q_0= n^a_{0}$, 
$\rho_0= n^a_{0}m$, and $p_0=0$ 
previously identified in Eqs.~(\ref{p.rho.sum}) (where we have already neglected the mean occupation number of antiparticles in the lowest energy level).

In practice, the critical temperature of condensation, $\beta_c$, is defined (implicitly) in terms of the cosmological charge density, $q$,
through the equation
\begin{equation}\label{eq.condition}
 q=q_{th}[\beta=\beta_c,\mu= m]\,.
\end{equation}
Here the subindex {\it th} makes reference to the fact that only the excited thermal modes, $k\neq 0$, have been included in this expression, $q_{\beta}=q_0+q_{th}$.
The right hand side in Eq.~(\ref{eq.condition}) codifies the maximum amount of charge $q$ per unit volume that, 
at a given temperature $\beta=\beta_c$, can be accommodated in the thermal modes once the
chemical potential has already reached its critical value at $\mu=m$. 
Below this temperature the chemical potential cannot grow anymore, and a macroscopic amount
of charge $q_0=q-q_{th}[\beta>\beta_c,\mu=m]$ must be accommodated in the ground state. %: the condensate emerges. 

In general, Eq.~(\ref{eq.condition}) should be evaluated numerically. However, this will not be necessary for the purposes of this paper.
As we will find next in Section~\ref{sec:BECDM}, for the case of a cosmological condensate
of DM particles thermal equilibrium must be broken during the relativistic, high charge density regime, 
where analytic expressions do exist.

The relativistic condensation of a gas of bosons in flat spacetime was considered for the first time in a seminal paper by Haber and Weldon~\cite{Haber}; 
see also Refs.~\cite{others,grether} for more recent works. Now we repeat this analysis for the case of 
a homogeneous and isotropic universe in a regime of slow expansion, $H\ll m$.
The functions $N^a_k k^2$ and $N^b_k k^2$ in Eqs.~(\ref{p.rho.sum}) are peaked around comoving wavenumbers $\beta a^{-1}k\sim 1$, and then only those values of 
$k$ contribute significantly to the integrals at different times in cosmic evolution.
In the relativistic regime, $\beta m\ll 1$, and to the lowest nonvanishing order in $m/(a^{-1}k)$, we can approximate the dispersion relation in 
Eq.~(\ref{cosmo.waves}) to $\omega_k=a^{-1}k$. Introducing this expression into Eqs.~(\ref{p.rho.sum}), we obtain 
\begin{subequations}\label{n,q,rho,o.rel}
\begin{eqnarray}
 n &=& (n^a_{0}+n^b_{0})+ \frac{1}{\pi^2\beta^3}\left[\textrm{Li}_3\left(e^{\beta\mu}\right)+\textrm{Li}_3\left(e^{-\beta\mu}\right) \right] \,,\\
 q &=& (n^a_{0}-n^b_{0})+ \frac{1}{\pi^2\beta^3}\left[\textrm{Li}_3\left(e^{\beta\mu}\right)-\textrm{Li}_3\left(e^{-\beta\mu}\right) \right] \,,
 \label{eq.q.rel}\\
 \rho &=& (n^a_{0}+n^b_{0})m+ \frac{3}{\pi^2\beta^4}\left[\textrm{Li}_4\left(e^{\beta\mu}\right)+\textrm{Li}_4\left(e^{-\beta\mu}\right) \right] \,,\label{rho.rel}\\
 p &=& \frac{1}{\pi^2\beta^4}\left[\textrm{Li}_4\left(e^{\beta\mu}\right)+\textrm{Li}_4\left(e^{-\beta\mu}\right) \right] \,. \label{p.rel} 
\end{eqnarray}
\end{subequations}
Here $\textrm{Li}_{\alpha}(x)$ denotes the polylogarithm function of order $\alpha$ and argument $x$. Some limiting values that 
will be of interest soon are given in Appendix~\ref{app.poly}. 
From the thermal nonzero-modes in Eq.~(\ref{eq.q.rel}) and the definition of the critical temperature of condensation 
in Eq.~(\ref{eq.condition}), we can read
\begin{subequations}\label{eqs.rel}
\begin{equation}\label{eq.temperature.rel}
 T_c=\sqrt{\frac{3q}{m}} \,,
\end{equation}
where we have made use of some of the expressions in Appendix~\ref{app.poly} in order to arrive to the final result. 
Note that the presence of particles as well as antiparticles is relevant at high temperatures,
and then the expression for the critical temperature of condensation should be modified with respect to that in standard textbooks on statistical 
mechanics, see e.g. Ref.~\cite{Huang}. The integral terms in Eqs.~(\ref{p.rho.sum}) only counts 
the excited thermal modes; for the configurations
below the the critical temperature of condensation a macroscopic amount of charge must be accommodated in the lowest energy zero-mode. 
We should then write $q=q_0+q_{th}[\beta,\mu=m]$, and the charge density in the
zero-mode can be evaluated from 
\begin{equation}\label{eq.particles.rel}
 q_0(T)= q\left[ 1 -\left(\frac{T}{T_c}\right)^2\right] \,,\quad \textrm{if}\quad T< T_c\,.
\end{equation}
\end{subequations}
As long as $q\gg m^3$ we can guarantee a relativistic temperature of condensation,
$T_c\gg m$, and then the approximations in Eqs.~(\ref{n,q,rho,o.rel}) are justified. 

Note that for a gas composed of particles with vanishing mass, the critical temperature
of condensation grows to infinity. This might seem to suggest that, in the zero-mass limit, 
any charge in the universe should be necessarily accommodated in the ground state. However, remember that the 
presentation in this paper is only valid as long as the expansion rate of the universe 
is not too large when compared to the characteristic time of oscillation of the zero-mode [this comes from the choice we made
in Eq.~(\ref{cosmo.waves}) for the mode-functions that solve Eqs.~(\ref{KG}) and~(\ref{normalization})], and only if the gas 
is in thermal equilibrium. In particular, we cannot extrapolate the expressions in this section to the massless case.
As we will discuss later in Section~\ref{sec:BECDM}, in order have a successful cosmological picture
any sensible DM candidate should be already decoupled from the thermal plasma at BBN, 
and then some caution is necessary even for non-zero mass particles; we will give some numbers soon.

The expressions for the energy density and pressure in the thermal gas are quite illuminating.
We can obtain these two quantities 
expanding Eqs.~(\ref{rho.rel}) and~(\ref{p.rel}) around $\beta\mu\ll 1$. After some algebra we get
\begin{equation}\label{eq.rho.p.rel}
 \rho= q_0(T) m+\frac{\pi^2T^4}{15}\,,\quad p=\frac{\pi^2T^4}{45}\,.
\end{equation}
Note that there are two different contributions to the expressions in Eq.~(\ref{eq.rho.p.rel}). 
On the one hand that coming from the particles in the condensate, $\rho= q_0(T) m$, $p=0$, that of course coincides with the contribution %the energy density and pressure 
associated to the order parameter in Eq.~(\ref{exp.zero.mode}).
The transition to a condensate is quite sharp: according to the expression in Eq.~(\ref{eq.particles.rel}), if the temperature 
is only one order of magnitude lower than the critical temperature of condensation, then 
$99\%$ of the charge will be already accommodated in the lowest energy level. As a consequence of this
we can think of $q_0(T)$ as a function that steeply transits from zero to the total charge density $q$ when the system approaches the critical 
temperature of condensation from above. In addition, and as a consequence of the high temperature in the gas, $\beta m\gg 1$, there is 
relativistic gas of thermal bosons, $\rho=\pi^2 T^4/15$, $p=\pi^2 T^4/45$, that appears even in the presence of a condensate when the temperature
of the universe is lower than the critical temperature. This thermal cloud will be essential for the analysis in the next section.

That the coherent excitation of the zero-mode appears only below a critical temperature of condensation might suggest that it is a late-time phenomena, 
characteristic of an old and cold universe. However, the critical temperature of condensation is not constant in cosmic history, and changes with 
the expansion of the universe as $T_c\sim 1/a^{3/2}$.
Then, the critical temperature of condensation decreases with the cosmological expansion faster than the temperature of a universe
in adiabatic expansion, where $T\sim 1/a$ (remember that it is the temperature of the CMB photons that determines the 
temperature of the gas, as long as they are in thermal equilibrium due to the self-annihilation of DM particles into the standard model sector).
Therefore, given an asymmetry between the number densities of particles and antiparticles, the condition $T\ll T_c$ is most easily satisfied in the {\em early} universe. 
Let us explore in more detail this scenario for the case of DM particles.

\section{The condensate as a viable candidate for dark matter}\label{sec:BECDM}

A potential CDM candidate requires, among other things, of an energy density with the right amount, $\rho(a=1)=\rho_{\textrm{CDM,now}}$, 
that is already decoupled from the thermal plasma at BBN and redshifts like a nonrelativistic component, $\rho\sim 1/a^3$, 
at matter-radiation equality. It is well known that particles with a mass less than about an eV  
that are produced thermally in the early universe cannot contribute significantly to the present matter content. 
This is a consequence of the Cowsik-McClelland bound~\cite{cowsik}, of which standard model neutrinos are a particular example.
Furthermore, the energy density in these low mass thermal particles redshifts with the cosmological expansion as radiation, $\rho\sim 1/a^4$, 
for a too large period exceeding the time of matter and radiation equality. They are therefore not a good prospect for CDM.
Let us see how this picture is modified in presence of an asymmetry.

For an asymmetric component the interactions with the thermal plasma can accommodate 
a significant amount of charge density in the ground state zero-mode. 
The particles in a condensate behave like standard CDM, at least at the level of the background universe,
and as long as the temperature is not close to the critical temperature of condensation so that the charge in the zero-mode does not change substantially 
with cosmic expansion, $q_0(T)\approx q$ (we already argued that the condensation is a very sharp process, and as we will find next for any sensible candidate
the temperature of decoupling is always well below the critical temperature of condensation). 
In order to see this note that, according to the identities in Eq.~(\ref{eq.rho.p.rel}), 
the particles in the condensate contribute to the energy density and pressure of the gas in the form $\rho=qm$, $p=0$.
The charge conservation within a comoving volume, $q\sim 1/a^3$, guarantees then the standard density evolution of a nonrelativistic component, 
$\rho\sim 1/a^3$. Note that since $\rho =qm$, the amount of CDM is now fixed by an asymmetry,
and it is not the result of the relics left dynamically at thermal decoupling (we will discuss about the relics left by 
a thermal condensate later in this section). There is no problem {\it a priori} with low mass candidates 
as long as they are not too light, $m>H_{d}$, so that the description in this paper applies and the previous simple arguments make sense.
Here $H_{d}$ is the value of the Hubble parameter at thermal decoupling. Since DM should be already decoupled at BBN, that imposes the constraint
\begin{equation}
m> H_{\textrm{BBN}}\sim  10^{-14}\,\textrm{eV}
\end{equation}
on the mass of the scalar particle. Later in Section~\ref{subsection.2} we will increase this number to $10^{-12}\,\textrm{eV}$ using
current bounds on the effective number of extra relativistic species.
Candidates lighter than this value require a more detailed analysis that we leave for another paper.
Once the particles are decoupled from the thermal plasma the occupation numbers get frozen and they just redshift with the 
volume of the universe, $n^{a,b}_k(a>a_d)=n^{a,b}_k(a=T_d)(a_d/a)^3$. This guarantees that the condensate does not evaporate if 
it did not before the time of decoupling, even if the temperature of the universe can grow above
the critical temperature of condensation at later times (actually the critical temperature is not even well defined if the gas is not in thermal equilibrium). 

Let us put some numbers to this. We can parametrize the DM asymmetry in terms of its ratio to the photon number density,
\begin{equation}\label{eq:eta_dm}
 \eta_{\textrm{CDM}}\equiv\frac{q_{\textrm{now}}}{n_{\gamma,\textrm{now}}} = 25.505(\Omega_{\textrm{CDM}}h^2) \left(\frac{\textrm{eV}}{m}\right)\,,
\end{equation}
where $q=\rho_{\textrm{CDM,now}}/(ma^3)$, $\Omega_{\textrm{CDM}}=\rho_{\textrm{CDM,now}}/\rho_{c,\textrm{now}}$, $\rho_{c}=3M_{\textrm{Pl}}^2 H^2/(8\pi)$, and 
$n_{\gamma}=2\zeta(3)T^3/\pi^2$. Today $H_{\textrm{now}}=100h\,\textrm{km/(s\,Mpc)}$, and $T_{\textrm{now}}=2.7255\pm 0.0006\,$K~\cite{fixsen}, with 
$M_{\textrm{Pl}}\approx 1.22\times 10^{19}\,\textrm{GeV}$ the Planck mass.
This assumes only one component of CDM, and also that all the %CDM 
particles in that component are in a condensed phase. We will relax these assumptions at the end of 
this section. Below we will find that for an asymmetric condensed DM component cosmological observations demand $m\lesssim 100\,\textrm{eV}$.
According to PLANCK $\Omega_{\textrm{CDM}}h^2= 0.1198\pm0.0015$ at 1$\sigma$ (68\%CL)~\cite{Planck}, and thus
for a successful candidate we need a mechanism that generates a fairly large DM asymmetry, $\eta_{\textrm{CDM}}\gtrsim10^{-2}$; compare this with e.g. the baryon-to-photon ratio, 
$\eta_{\textrm{bar}}\sim 10^{-10}$~\cite{PDG}. We are not going to explore this mechanism here, but again let us stress  
that it is this large value of the asymmetry what makes it possible that light particles behave like a nonrelativistic component in spite of being 
in thermal equilibrium in the early universe.

\begin{figure}[!t]
 \includegraphics[width=0.8\textwidth]{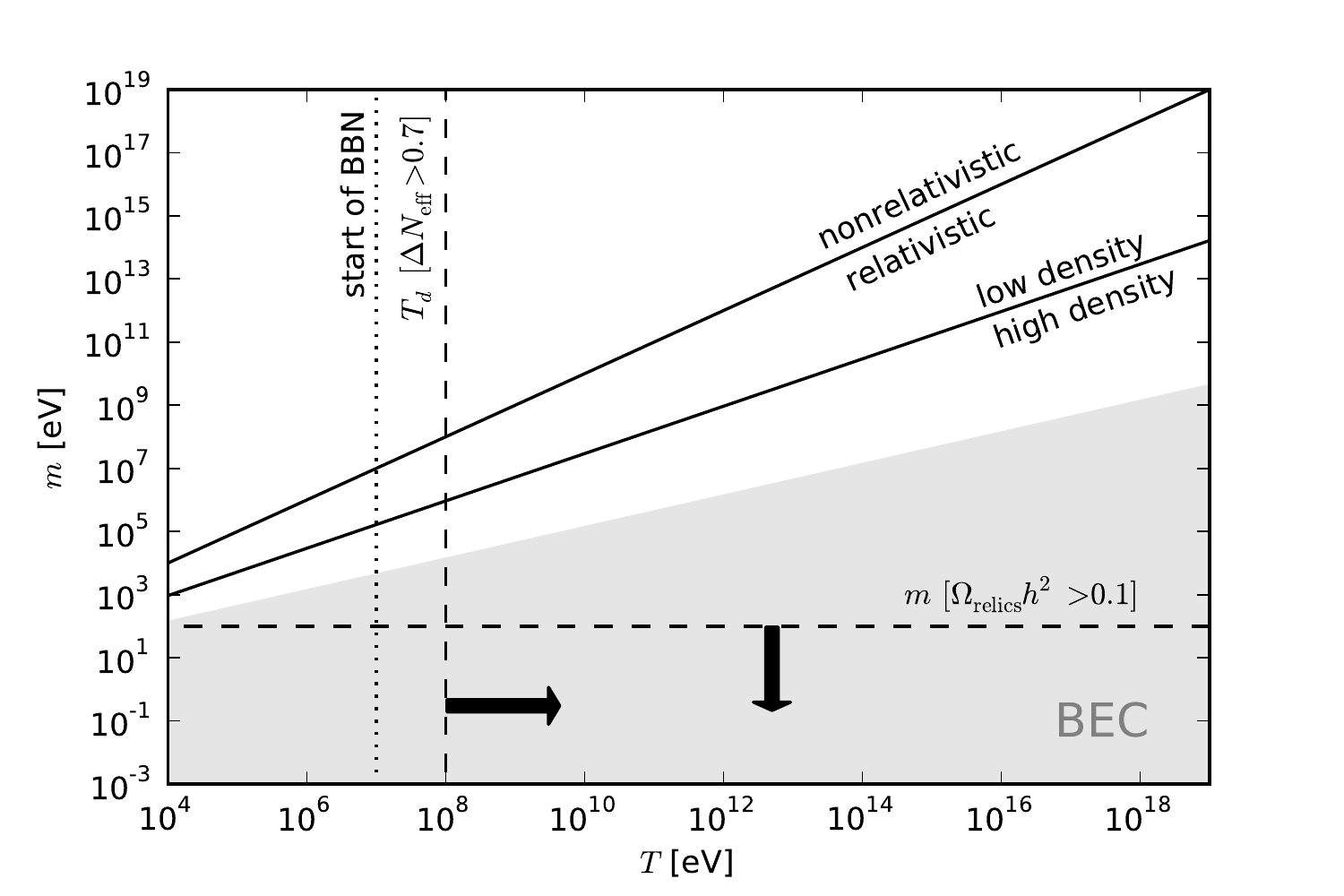} 
 \caption{
 Cosmological evolution of a gas of bosons with a non-zero charge density 
 as a function of the temperature of the CMB photons $T$ for different masses of the scalar particle $m$.
 The shaded region indicates the presence of a condensate, $T<T_c$. The transition from a relativistic to a nonrelativistic gas occurs
 at $T\sim m$ in Eq.~(\ref{T(z)}), whereas the transition from a universe with a
 high-charge-density to one with a low-charge-density takes place at $q \sim m^3$ in Eq.~(\ref{q(z)}).
 Here we have considered that the condensate contains the totality of the CDM, and that it is in equilibrium with the thermal plasma. 
 %although the constraints remain valid even if it contributes only to a fraction of the total CDM in the universe.
 Note that the particles that could conform a cosmological DM condensate should be necessarily much lighter and weakly interacting than usually expected for standard WIMPs, 
 and that for the allowed range of masses the temperature of decoupling $T_d$ is always well below the critical temperature of condensation 
 (see Sections~\ref{subsection.1} and~\ref{subsection.2} for further details). The dotted line represent the beginning of BBN and are just for orientation.
 This figure is {\it not} applicable to candidates lighter than around $10^{-12}\,\textrm{eV}$ since they cannot have reached the regime of slow expansion,
 $H\ll m$, at decoupling.
 }
 \label{fig:cosmological evolution}
\end{figure}

As long as the scalar particles are in thermal equilibrium with the constituents of the standard model of particle physics, they will share a common temperature.
Using the relation $T\sim 1/a\sim (1+z)$ for a universe in adiabatic expansion, we can determine this temperature as a function of time in terms of the temperature 
in the CMB photons today,\footnote{Properly speaking, from the conservation of the entropy in a universe in adiabatic expansion, $s(T)a^3=s(T_{\textrm{now}})$,
we should write $T=[g_{*S}(T_{\textrm{now}})/g_{*S}(T)]^{1/3}T_{\textrm{now}}(1+z)$, where the function $g_{*S}(T)$ parameterizes the effective number of relativistic 
species contributing to the entropy density $s$ at temperature $T$, see Eqs.~(\ref{eq:entropy.density}) and~(\ref{eq:def.g}) below  
for details. Then, the identity in Eq.~(\ref{T(z)}) is only satisfied after the electron-positron annihilation at $T\sim 0.5\,\textrm{MeV}$,
once the function $g_{*S}(T)$ gets fixed to the current value $g_{*S}(T_{\textrm{now}})$.
However, in practice and for the model we are considering in this paper
$0.3\lesssim[g_{*S}(T_{\textrm{now}})/g_{*S}(T)]^{1/3}\le 1$, 
and this factor is not going to be relevant for what we want to show here. Furthermore, this extra factor can only reduce the actual value 
of the temperature with respect to 
that in Eq.~(\ref{T(z)}), and then it contributes favorably to the appearance of a condensate.}
\begin{equation}\label{T(z)}
 T(z)=8.617\times 10^{-5}\left(\frac{T_{\textrm{now}}}{\textrm{K}}\right)(1+z)\,\textrm{eV}\,.
\end{equation}
On the other hand, from Eq.~(\ref{eq:eta_dm}) and the relation $q\sim 1/a^3\sim(1+z)^3$ for the charge conservation,
we can determine the charge density necessary to match the totality of the CDM, 
also as a function of time,
\begin{equation}\label{q(z)}
 q(z)=\,8.127\times 10^{-11}\left(\frac{\textrm{eV}}{m}\right)(\Omega_{\textrm{CDM}}h^2)(1+z)^3\,\textrm{eV}^3\,.
\end{equation}
Combining Eqs.~(\ref{T(z)}) and~(\ref{q(z)}) we can infer when $i)$~the transition at $T\sim m$ from a relativistic to a nonrelativistic gas of bosons occurs; 
$ii)$~the transition at $q\sim m^3$ from a universe with a high charge density to another  with a low charge density takes place; and $iii)$~the
transition at $T= T_c$ from a condensate to a phase with no condensate happens. All these regimes are illustrated in Figure~\ref{fig:cosmological evolution}.

We can look at this figure as representing, for different masses of the scalar particle, the 
phases of a cosmological gas of bosons with a charge asymmetry that matches the totality of the CDM, whenever the bosons are in equilibrium with the thermal plasma.
Note that, for a fixed value of the mass of the scalar particle, the condensate is present until the  
temperature of the universe reaches the critical temperature of condensation, when the condensate in principle disappears: 
if we move from the right to the left in Figure~\ref{fig:cosmological evolution} following horizontal lines we eventually leave the shaded region.
That means that we need to break thermal equilibrium before reaching that point, so that the amount of charge in the condensate freezes out. 
Furthermore, in order to do not affect the abundance of light elements predicted in the standard cosmological scenario,
the temperature of decoupling should be at least as high as the temperature in the universe at the beginning of BBN, $T\sim 10\,\textrm{MeV}$. 
Then, we can easily appreciate from Figure~\ref{fig:cosmological evolution} that the condensed phase 
is necessarily in the relativistic, high density regime, i.e. prior to the beginning of BBN the line that divides the relativistic and the nonrelativistic
regimes is always above the shaded region. That implies that, according to the expressions in Eq.~(\ref{eq.rho.p.rel}), 
apart from the condensate there will be a thermal cloud of relativistic particles. This cloud also freezes at the time of decoupling, producing (hot) thermal 
relics that will be around in the late-time universe and could affect the observations. This is a consequence of the thermal origin of the condensate.

Let us explore in more detail the cosmological consequences of these thermal relics. Since the DM particles decoupled being hot, we can easily compute their temperature 
(which is only well defined as long as the particles are relativistic, and moreover does not necessarily coincide with the temperature of the CMB photons), 
and then their abundance in the present universe. In a universe in adiabatic 
expansion the comoving entropy $S$ is conserved, and then $S=s(T)a^3=s(T_{\textrm{now}})$. Here 
\begin{equation}\label{eq:entropy.density}
 s(T)= \frac{2\pi^2}{45}g_{*S}(T)T^3
\end{equation}
is the entropy per unit volume, and the function
\begin{equation}\label{eq:def.g}
 g_{*S}(T) = \sum_{i\textrm{ bosons}} g_i\left(\frac{T_i}{T}\right)^3 + \frac{7}{8} \sum_{i\textrm{ fermions}} g_i\left(\frac{T_i}{T}\right)^3
\end{equation}
counts the total number of effectively massless degrees of freedom contributing to the entropy density at temperature $T$, i.e.
those that are in thermal equilibrium at a temperature $T_i(T) \gg m_i$, or that are not in
thermal equilibrium anymore but decoupled and froze out when they were still relativistic. Here $g_i$ is the number of spin states of the particle and 
antiparticle $i$, and the factor of $7/8$ in the second term accounts the difference between fermions and bosons.
%Note that, in general, $T_i \leq T$ if the component $i$ is already decoupled from the thermal plasma. 
From the equation of entropy conservation, $S=S_{\textrm{SM}}+S_{\textrm{relics}}$, we obtain 
\begin{equation}\label{eq:temperature.1}
 T(a) = \left(\frac{g^{\textrm{SM}}_{*S}(T_d)}{g^{\textrm{SM}}_{*S}(T)}\right)^{1/3}\left(\frac{a_d}{a}\right)T_d  
\end{equation}
for the temperature of the CMB photons as a function of time, whereas 
\begin{equation}\label{eq:temperature.2}
 T_{\textrm{relics}}(a) = \left(\frac{a_d}{a}\right)T_d 
\end{equation}
for the temperature in thermal relics also as a function of time. Here $T_d$ and $a_d$ are the temperature and the scale factor at which DM decoupled from the primordial plasma, 
and we have used the fact that the particles in the standard model and the thermal relics do not interact after they decoupled, so that the comoving
entropies $S_{\textrm{SM}}$ and $S_{\textrm{relics}}$ are conserved independently. The two expressions above are valid only after DM decoupling, 
with $T_{\textrm{DM}}(T>T_d)=T$ before that time. 
Here the superscript SM in $g_*^{\textrm{SM}}(T)$ makes reference to the fact that this function only takes into account the
relativistic degrees of freedom in the standard model.
For the standard model of particles
physics this quantity is tabulated and can be found in the literature, see e.g. Fig. 3.5 in Ref.~\cite{KolbTurner}.
Note that $g^{\textrm{SM}}_{*S}(T)\le 106.75$, 
and today $g^{\textrm{SM}}_{*S}(T_{\textrm{now}})=3.909$, where photons, $g_{\gamma}=2$, $T_{\gamma}=T$, and neutrinos, 
$g_{\nu}=6$, $T_{\nu}=(4/11)^{1/3}T$, are not in thermal equilibrium but decoupled when
they were still relativistic. This upper bound on $g^{\textrm{SM}}_{*S}(T)$ will be useful soon.

Combining the two identities in Eqs.~(\ref{eq:temperature.1}) and~(\ref{eq:temperature.2}), we obtain
\begin{equation}\label{eq:Tdm/T}
 \frac{T_{\textrm{relics}}}{T} = \left(\frac{g^{\textrm{SM}}_{*S}(T)}{g^{\textrm{SM}}_{*S}(T_d)}\right)^{1/3}\,,
\end{equation}
where again this expression is only valid at temperatures $T<T_d$. Note that if after DM decoupling a standard model particle that is in thermal equilibrium becomes nonrelativistic and annihilates 
(decreasing the value of the effective number of relativistic species $g^{\textrm{SM}}_{*S}$), its entropy is transferred to the other particles that are still 
in equilibrium, but not to the scalar bosons, ``reheating" the CMB. That is the reason for which, in general, $T_{\textrm{relics}}\neq T$. 

The expression in Eq.~(\ref{eq:Tdm/T}) is valid only as long as the DM particles are relativistic. Once they go nonrelativistic the thermal spectrum breaks down, 
and in general we cannot define a proper temperature for the boson relics. However, we can still write
\begin{equation}\label{eq:ratio.2}
 \frac{n_{\textrm{relics}}}{n_{\gamma}} = \frac{g^{\textrm{SM}}_{*S}(T)}{g^{\textrm{SM}}_{*S}(T_d)}\,.
\end{equation}
This expression comes from the conservation of the comoving number density of thermal relics after decoupling, $N=n_{\textrm{relics}}(T_d)/s(T_d)=n_{\textrm{relics}}(T)/s(T)$,
where we have used the identities $n_{\textrm{relics}}(T)=n_{\gamma}(T)=2\zeta(3)T^3/\pi^2$ and $g_{*S}(T)/g_{*S}(T_d)=g^{\textrm{SM}}_{*S}(T)/g^{\textrm{SM}}_{*S}(T_d)$.
Once again the number of thermal relics per photon decreases whenever a particle species self-annihilates and disappears heating up the CMB photons. Incidentally, for the model we are 
considering in this paper the two ratios in Eqs.~(\ref{eq:Tdm/T}) and~(\ref{eq:ratio.2}) have not changed since the electron-positron annihilation at $T\sim 0.5\,\textrm{MeV}$.

Note that since we are dealing with hot thermal relics they are almost as abundant today as the photons in the CMB, see Eq.~(\ref{eq:ratio.2}) above, and they can affect the 
cosmological observations in two different ways: if they are too ``heavy'', $m\gtrsim 10^{-4}\,$eV, they could comprise too much DM; if on the contrary they are too 
``light'', $m\lesssim 10^{6}\,$eV, they could potentially spoil the success of BBN. Let us now discuss these two constraints in turn.

\subsubsection{Dark matter particles of masses $m\gtrsim 10^{-4}\,\textrm{eV}$}\label{subsection.1}

If the mass of the scalar particle is too large, $m\gtrsim 10^{-4}\,$eV, the thermal relics will be nonrelativistic today, 
$m>T_{\textrm{relics,now}}\le T_{\textrm{now}}\sim 10^{-4}\,$eV, and they will contribute to the mass density of the universe according to
\begin{equation}
 \Omega_{\textrm{relics}}h^2 = 0.153\frac{1}{g^{\textrm{SM}}_{*S}(T_d)}\left(\frac{m}{\textrm{eV}}\right)\,,
\end{equation}
where $\Omega_{\textrm{relics}}=\rho_{\textrm{relics,now}}/\rho_{c,\textrm{now}}$, $\rho_{\textrm{relics}}=n_{\textrm{relics}}m$, 
$n_{\textrm{relics}}=2\zeta(3)T_{\textrm{relics}}^3/\pi^2$,  $\rho_{c}=3M_{\textrm{Pl}}^2 H^2/(8\pi)$, and today $g^{\textrm{SM}}_{*S}(T_{\textrm{now}})=3.909$. 
Note that the lighter and more weakly interacting [i.e. the larger $g^{\textrm{SM}}_{*S}(T_d)$] the scalar particles are, the less they will contribute to the matter content.
Even in the best case scenario in which $g^{\textrm{SM}}_{*S}(T_d)= 106.75$, thermal relics heavier than  $71.546\,$eV
will contribute more than $\Omega_{\textrm{relecis}}h^2\sim 0.1$ to the energy density of the present universe, and we can then safely exclude these values 
for the mass of the scalar bosons; see Figure~\ref{fig:cosmological evolution} for details.
Compare this with the Lee-Weinberg bound for standard cold WIMPs, where $m\gtrsim 10\,\textrm{GeV}$~\cite{Lee-Weinberg} 
(particles of mass $10\,\textrm{eV}\lesssim m\lesssim 10\,\textrm{GeV}$ that are produced thermally and interact mainly through the weak nuclear force
would represent too high a DM-to-photon ratio, whereas particles of mass $m\lesssim 10\,\textrm{eV}$ would contribute 
as hot DM and present problems with e.g. structure formation).

\subsubsection{Dark matter particles of masses $m\lesssim 10^{6}\,$eV}\label{subsection.2}

If on the contrary the mass of the scalar particle is too low, $m\lesssim 10^{6}\,$eV, the thermal relics will be relativistic at the time of BBN, 
$T_{\textrm{BBN}}\sim 10-0.1\,$MeV, and they will contribute to the effective number of extra neutrino species %at that time 
in the form
\begin{equation}
 \Delta N_{\textrm{eff}}=27.114\frac{1}{[g^{\textrm{SM}}_{*S}(T_d)]^{4/3}}\,,
\end{equation}
where $\Delta N_{\textrm{eff}}$ is defined through
\begin{equation}
 \rho_{\textrm{rad}}=\frac{\pi^2}{30}g_*(T)T^4=\left[1+\frac{7}{8}\left(3.046+\Delta N_{\textrm{eff}}\right)\left(\frac{4}{11}\right)^{4/3}\right]\rho_{\gamma} \,.
\end{equation}
Here $\rho_{\textrm{rad}}$ denotes the energy density in radiation, and $\rho_{\gamma}=(\pi^2/15)T^4$ 
that in photons. Note that, actually, for the whole range of masses that survives the previous constraint in Section~\ref{subsection.1}, $m\lesssim 100\,\textrm{eV}$, 
the particles are relativistic at BBN, and then the discussion in this section is general.
As usual $N_{\textrm{eff}}^{\textrm{SM}}=3.046$ represents the effective number of neutrinos in the standard model of particle physics~\cite{mangano},
and $T_{\nu}/T=(4/11)^{1/3}$ [$T_{\textrm{relics}}/T=(3.909/g^{\textrm{SM}}_{*S}(T_d))^{1/3}$] is the ratio of the temperature in neutrinos [thermal relics] to that in 
photons after the electron-positron annihilation. Once again, the more weakly interacting the scalar particles are, the less they will contribute to the energy budget, 
i.e. to the effective number of neutrino species. However, since we are now dealing with relativistic relics, and contrary to what happens in Section~\ref{subsection.1}, 
this contribution is not sensitive to the actual mass of the particles. 

An analysis by Cyburt {\it et al} combining observations of the primordial $^4$He mass fraction, light element abundances, and baryon-to-photon ratio 
($Y_p$+D/H$_A$+$\eta_{\textrm{CMB}}$) results in an upper bound of $\Delta N_{\textrm{eff}}^{\textrm{BBN}}<0.804$ 
at 68\%CL to the extra neutrino species at BBN~\cite{Cyburt}, and then in a lower bound of $g^{\textrm{SM}}_{*S}(T_d)>13.994$
to the effective number of relativistic species contributing to the entropy density also at BBN. 
Using the latest data provided by the PLANCK collaboration in a joint CMB+BBN analysis 
(D+{\it Planck} TT,TE,EE+lowP), $\Delta N_{\textrm{eff}}^{\textrm{CMB}}<0.234$~\cite{Planck},
we can improve this bound to $g^{\textrm{SM}}_{*S}(T_d)>35.317$, now at 95\%CL. 
Note however that this latter constraint applies only for DM bosons of mass less than about an eV,
which are still relativistic at recombination. 
These values of $g^{\textrm{SM}}_{*S}$ lie around the scale of the QCD phase transition, $\Lambda_{\textrm{QCD}}\sim 200\,$MeV, where the 
baryons and mesons are formed and the number of massless degrees of freedom drops drastically from $g^{\textrm{SM}}_{*S}\sim 60$
to $g^{\textrm{SM}}_{*S}\sim 20$. We can then conclude that $T_d\gtrsim\Lambda_{\textrm{QCD}}$ for an asymmetric condensed DM candidate.

Two comments are in order here. First, since the number of massless degrees of freedom in the standard 
model is constrained to $g^{\textrm{SM}}_{*S}(T)\le 106.75$, then there is a lower bound for the number of extra neutrino species: 
$\Delta N_{\textrm{eff}}>0.054$. 
This value is around the sensitivity of $\pm 0.04$ on $N_{\textrm{eff}}$ expected from future measurements of the polarization of the CMB~\cite{brust}, 
and then the cosmological observations could test the viability of this model soon. 
Second, the lower bound of $T_d\gtrsim 200\,\textrm{MeV}$ on the temperature of freeze out
is related to the self-annihilation, DM+DM$\rightleftharpoons$SM+SM, and elastic scattering, DM+SM$\rightleftharpoons$DM+SM, 
cross sections associated to the processes that maintained the condensate in chemical and kinetic equilibrium 
with the thermal plasma in the early universe, respectively, and then to the strength of the interactions between DM and standard model particles.
As a consequence of this high value of the temperature of DM decoupling the interaction of the bosons with the standard model particles should be very weak,
even weaker than in the case of ordinary WIMPs, where e.g. $T_d\sim 1\,$MeV for neutrinos.\footnote{For
the case of standard cold 
thermal relics the temperature of decoupling can be also as high as $T_d\sim 100\textrm{MeV}$, but 
there this is a consequence of the rarefaction of the gas due to the unbalanced self-annihilation of DM into standard model particles,
and not of the weakness of the interactions.} Note also in Figure~\ref{fig:cosmological evolution} that 
for the allowed range of masses of the scalar particle the critical temperature of condensation
is always above the temperature at which thermal equilibrium broke down, and also above the temperature of BBN. 
This would guarantee the survival of the coherent state to thermal evaporation, however, 
this point deserves a more detailed analysis of the kinetic and chemical decoupling in terms of the Boltzmann equation in the presence of a condensate.

Let us come back to the general picture. One could imagine scenarios in which the condensate would represent only a fraction of the total CDM in the universe, the remainder being
in the form of standard thermal relics or any other candidate to DM. 
In this scenario the lines in Figure~\ref{fig:cosmological evolution} dividing the high-density/low-density, 
and the condensed/non-condensed, phases move down (note that they are sensitive to the 
asymmetry of the universe, and in particular they disappear if $q=0$), 
but the other information in the
figure remains unaltered. This does not affect, in particular, the 
conclusion that the thermal relics accompanying the condensate are hot, nor the upper bound we found previously in Section~\ref{subsection.1} for the mass of the scalar particles.  
Thermal relics of mass  $m \lesssim 100\,$eV would contribute to the energy density of the universe as hot DM, and
they cannot represent a significant part of the matter content.\footnote{Hot thermal relics free-stream in the early universe. 
This avoids gravitational instability and induces a cutoff in the mass power spectrum at the scale 
$M_J\sim M_{\textrm{Pl}}m^{-2}\sim  (10\,\textrm{keV}/m)^2\times 10^8 M_{\odot}$~\cite{bond}. In order to not erase dwarf galaxies, $M_{\textrm{dwarfs}}\sim
10^9M_{\odot}$, and heavier structures we should demand $m\gtrsim10\,\textrm{keV}$.} Then, if there are no more particles in the universe 
apart from the DM boson and the constituents of the standard model, we can conclude that the totality of the CDM should be necessarily in a light condensate, 
or in the form of heavy thermal relics, but we cannot have a combination of both. This last statement could be avoided, of course, if we had another component of DM
(apart from the boson particles that conform the condensate) in the form of e.g. WIMPs, but the previous constraints on the scalar particles still apply.

To summarize, the scalar DM scenario we presented in this paper has necessarily a condensed and a thermal component. 
The condensed component is essentially determined by a charge asymmetry, which must be
substantial to provide enough CDM. Lest the thermal component represent {\em too much} DM (or the right amount, but with wiping out too much small-scale 
structure), it must be lighter than about a hundred of eV's, and then the DM boson must decouple at high temperature so that it does not add a light degree of freedom during 
BBN. These constraints remain even if the scalar DM particles comprise just a significant part, rather than all, of the observed CDM. 

$\vspace{0.01cm}$

\centerline{*\quad *\quad *}

$\vspace{0.01cm}$

Since early decoupling is crucial, it is worth some discussion here. The question of decoupling is intimately related with the question of what led the system into thermal 
equilibrium. For a standard thermal candidate 
it is usually argued that gravity is not sufficient, and there should be an additional interaction, usually the weak nuclear force, that couples DM to the particles 
in the standard model. Here the argument goes as follows~\cite{KolbTurner}: For gravity the interaction rate scales like 
$\Gamma_{\textrm{grav}}\sim n_{\textrm{DM}}\langle\sigma v\rangle\sim T^5/M_{\textrm{Pl}}^4$, where 
$n_{\textrm{DM}}\sim T^3$ and $\langle\sigma v\rangle\sim T^2/M_{\textrm{Pl}}^4$. On the other hand, during the radiation dominated era the 
Hubble parameter is proportional to $H\sim T^2/M_{\textrm{Pl}}$, and then decoupling, 
$\Gamma_{\textrm{grav}}\sim H$, occurs at $T_\textrm{dec}\sim M_{\textrm{Pl}}$. We cannot trust these expressions at the Planck scale, 
and then we conclude that the gravitational interaction alone could not have brought the early universe into  
thermal equilibrium.

However, for the case of an asymmetric DM component the number of particles scales with 
$n_{\textrm{DM}}\sim T^3+ q \sim (1+\eta_{\textrm{CDM}})T^3$, 
and we can then conclude that $T_d\sim (1+\eta_{\textrm{CDM}})^{-1/3}M_{\textrm{Pl}}$. For ultralight candidates, $\eta_{\textrm{CDM}}\gg 1$, the interaction rate increases 
with respect to the standard scenario and decoupling is expected to take place well after the Planck era, where we can now rely on these expressions.
This is of course only a very naive picture; a more accurate treatment would employ the Boltzmann equation~\cite{gondolo.boltz} 
to treat the decoupling process in the presence of a condensate. 
We leave this analysis for a future work.

\section{Conclusions}\label{sec:discussion}

In this paper we have considered the possibility that much or all of the CDM in the universe results from an asymmetry between the number densities 
of particles and antiparticles, rather by the abundance of thermal relics frozen out at the time of thermal decoupling.
We have identified three conditions that, if satisfied, guarantee the appearance of a thermal BEC during the early stages of the universe: 
$i)$~DM is described in terms of a boson field with a conserved current;
$ii)$~There is a nonvanishing total boson charge in the universe;  
and $iii)$~The boson was in thermal (kinetic and chemical) equilibrium with the particles in the standard model. 
If, in addition, $iv)$~Thermal decoupling took place at an energy scale below the critical temperature of condensation,
then the macroscopic excitation of the ground state freezes out and could survive the ensuing cosmological evolution until now. 
Note that for the purposes of this paper we did not need to develop any boson candidate, not even specify whether 
they are fundamental, or just an effective low energy description, e.g. made of pairs of fermions.
We have estimated the charge asymmetry required to account for the totality of the 
CDM using the condensate, but do not address here how this large asymmetry would have arisen.

According to our findings there is an upper limit on the mass of the DM particles in a thermal BEC.
This is because in this model the macroscopic excitation of the zero-mode has a relativistic origin and then, apart from the condensate, there will be  
thermal relics in an abundance close to that of the photons in the CMB. These particles must then have masses lower than around $100\,$eV 
so that their density is below the current cosmological constraints (compare this with the unitarity bound $m\lesssim 100\,\textrm{TeV}$ one finds
in the case of ordinary WIMPs~\cite{unitarity}).
The existence of thermal relics could also alter the success of BBN. In order to not contribute in excess to the effective number of neutrino species 
the relics should be cold enough, that is, they should decouple early on in the history of the universe.
Using the latest BBN and BBN+CMB observational data we obtain that the DM bosons should be already decoupled at the time of the QCD phase transition, 
when the temperature of the universe was of the order of $200\,$MeV. 
As a consequence the interaction between the DM particles that conform the condensate and those in the standard model should be even weaker than 
the weak nuclear force. 
We speculate about the possibility that gravity alone could be the responsible of the early thermal equilibrium.
Incidentally, there is a lower bound to the excess of neutrino species predicted by an asymmetric condensed DM component, $\Delta N_{\textrm{eff}}>0.054$.
This value could be resolved in the next future using CMB data, and then test the viability of the model.
 
Let us emphasize that, in spite of the early thermal equilibrium, the particles in a DM condensate could be as light as $m\sim 10^{-12}\,$eV 
and behave like a nonrelativistic component in terms of their background ``equation of state.''
(For candidates of masses lower than this value the decoupling from the thermal plasma necessarily occurs during  
the regime of fast expansion, $H \gg m$, and then a more elaborate
analysis is necessary to treat the phenomenology of ultralight particles.) 
We usually tend to think of thermal cold candidates as particles with large masses, as happens in e.g. the standard
{\it symmetric} cosmological scenario. However, if DM is described in terms of a boson field, there is another possibility 
to cool the particles in the universe: giving them a large asymmetry at the expense of low masses.

Note that even though the input of the asymmetric condensed DM scenario is very similar to that of ordinary WIMPs (i.e. a thermal quantum field theory),
the output differs drastically: whereas at the effective level we can think of WIMPs as a collection of classical {\it particles}, the condensate is 
described in terms of a classical {\it field} theory. This can affect, for instance, the process of structure formation at small scales, 
leading to possible interesting observational consequences in the late-time universe~\cite{fuzzy,Tom,jose, miguel, Tom2,marsh2}.

The analysis in this paper complements the standard thermal scenario of a boson DM candidate (without an asymmetry,
or with an asymmetry much lower than the value reported in Eq.~(\ref{eq:eta_dm}), so that the quantum Bose statistics is blurred by the classical Maxwellian
distribution for most of the history of the universe), and indicates that a boson particle that was in thermal equilibrium with the constituents of the standard
model of particle physics should be in a condensate, or in the form of thermal relics, but we cannot
have a combination of both aspects contributing significantly to the matter content today.

%%%%%%%%%%%%%%%%%%%%%%%%%%%
%%%   ACKNOWLEDGMENTS   %%%
%%%%%%%%%%%%%%%%%%%%%%%%%%%

\acknowledgments

We are grateful to Patrick Draper, Alma Gonzalez-Morales, David Marsh, Chanda Prescod-Weinstein, Stefano Profumo, Tanja Rindler-Daller and Luis Ure\~na-Lopez for useful comments and discussions.
AA and ADT are supported in part by Grant No.~FQXi-1301 from the Foundational Questions
Institute (FQXi).  ADT is also supported by CONACyT Mexico under Grants No.~182445 and No.~167335.

\appendix

\section{Some properties of the polylogarithm functions}\label{app.poly}

In this appendix we present some useful limiting values of the polylogarithm functions. They are useful when evaluating the expressions in 
Eqs.~(\ref{n,q,rho,o.rel}) in the main text.
\begin{subequations}
 \begin{eqnarray}
  \textrm{Li}_{3}(x\to 1) &=& \zeta(3)+\frac{\pi^2}{6}(x-1)+\ldots\,,\\
  \textrm{Li}_{4}(x\to 1) &=& \frac{\pi^4}{90}+\zeta(3)(x-1)+\ldots\,.
 \end{eqnarray}
\end{subequations}
Here $\zeta(x)$ is the Riemann zeta function, with $\zeta(3)\approx 1.2021$.

%%%%%%%%%%%%%%%%%%%%%%
%%%   REFERENCES   %%%
%%%%%%%%%%%%%%%%%%%%%%

%%%%%%%%%%%%%%%
%%%   END   %%%
%%%%%%%%%%%%%%%                 

\end{document}